\documentclass{elsart}
\usepackage{amssymb}
\usepackage{amsmath}
\usepackage{graphics}

\usepackage{epsfig}
\usepackage{amssymb}

\def\bfg #1{{\mbox{\boldmath $#1$}}}

\begin{document}

\begin{frontmatter}

\title{
Spin observables in the $NN\to  Y\Theta^+$ reaction at the threshold
and quantum numbers of the $\Theta^+$ pentaquark
}

\author{Yu.N.~Uzikov}
\footnote{E-mail address: uzikov@nusun.jinr.ru
}

\address{Joint Institute for Nuclear Research, LNP, 141980 Dubna,
 Moscow Region, Russia}


\begin{abstract}
{ General formulae  for the spin-spin correlation parameters
 $C_{i,j}$ and spin-transfer coefficients $K_i^j$ are derived for 
 the reaction $N\,N\to Y\,\Theta^+$ at the threshold for an 
 arbitrary spin of the pentaquark $\Theta^+$.
 It is shown that a measurement of the
 sign of $C_{y,y}$ or an observation of the non-zero polarization 
 transfer  from the nucleon to the hyperon $Y$ allow one to
 determine the P-parity of the $\Theta^+$ unambiguously and
 independently on the spin of the $\Theta^+$. Measurement of these
 spin observables in the  both $pp$- and $pn$- channels of this reaction
 determines also the isospin of the $\Theta^+$.}
\end{abstract}  

\begin{keyword}

Pentaquark, strangeness, spin observables

\begin{PACS}
13.88.+e, 13.75.Cs, 14.20.-c\\[1ex]
\end{PACS}
\end{keyword}

\end{frontmatter}
\baselineskip 4ex
\newpage

\section{Introduction}

  Experimental indications
 \cite{Nakano,Barmin,Stepanyan,Kubarovsky,Elsa,hermes,SVD}
 to  existence of an exotic barion 
 with the strangeness $S=+1$, called as the $\Theta^+(1540)$,
 which presumably consists of five constituent quarks, stimulated many
 theoretical works.  An important task now is an experimental
  determination of the quantum numbers of the $\Theta^+$.  
 Model independent methods for determination of the P-parity of the
 pentaquark $\Theta^+$ in the reaction $NN\to Y\,\Theta^+$ were
 suggested in Refs. \cite{thomas,hanhart,uzikov,rekalo2}. These
 methods are based on such   general properties of the reaction 
  amplitude  as  angular momentum and P-parity conservation and 
  on the generalized Pauli principle for  nucleons. 
 It was shown that
 the sign of the spin-spin correlation parameter $C_{y,y}$ 
 determines unambiguously the P-parity of the
 $\Theta^+$, $\pi_\Theta$,
 in the reaction $pp\to\Sigma^+\Theta^+$ \cite{hanhart}.
 Another strong
 correlation between  $C_{y,y}$ and $\pi_\Theta$
 is valid also for the $pn\to \Lambda^0 \Theta^+$ reaction 
 \cite{uzikov,rekalo2} if the isospin of the $\Theta^+$ 
 equals to zero.
 Furthermore,
 measurement of the spin transfer coefficients $K_y^y=K_x^x$ or
 $K_z^z$
 of these reactions also allow to
 determine the P-parity unambiguously  \cite{uzikov,rekalo2}.
 A measurement of the polarization transfer from the initial nucleon to the
 hyperon  in the reaction $NN\to Y\Theta^+$ can be perfomed 
 by a single spin experiment with
 polarized beam or target, because the polarization of the hyperon
 can be measured via its weak decay. 
 However, the results of Refs. \cite{thomas,hanhart,uzikov,rekalo2}
 are based on the assumption that the spin of the $\Theta^+$ 
 is equal to $\frac{1}{2}$. Up to now the spin of the $\Theta^+$
 is  not known, as well as the P-parity and isospin, and within some 
 models
 its value can be $\frac{3}{2}$. In this work we derive  formulae
 for  the spin
 observables of the reaction $NN\to Y\,\Theta^+$ at the threshold
 for general case of
  arbitrary spin of the $\Theta^+$.  Analysis is based on common
 properties of the reaction amplitude and 
 the standard  method of the spin-tensor operators \cite{bilenky}.
 We derive also a full spin structure of the cross section of 
 this reaction  for the case  of the  spin-$\frac{1}{2}$  particles
 taking into account  all polarizations  in the initial an
 final states.
\section{Formalism}

 Assuming  dominance of the S-wave in the
 relative motion in the final system,
 the most general expression for the amplitude of the binary reaction
 $1+2 \to 3+4$ at the threshold  can be written as \cite{GW}
\begin{equation}
\label{tfi}
T_{\mu_1\,\mu_2}^{\mu_3\,\mu_4}=\sum_{J\,M \,\atop
 S\,M_S\,L\,m}
(j_1\mu_1\,j_2 \mu_2|SM_S)
(j_3\mu_3\, j_4\, \mu_4|J\,M)(S\,M_S\,L\,m|J\,M)
Y_{Lm}({\hat {\bf k}}) a_J^{LS}.
\end{equation}
 Here $j_i$ and $\mu_i$ are the spin of the i-th particle and its  
 z-projection, $J$ and $M$ are the total angular
 momentum and its z-projection; 
 $S$ and $L$  are the  spin  and orbital momentum of the
  initial system, respectively, and $M_S$ and $m$ are the
 corresponding z-projections. An information on the reaction dynamics is
 contained in  the complex  amplitudes 
 $a_J^{L\,S}$. The sum over $J$ in Eq. (\ref{tfi}) is restricted
  by the conditions $J=j_3+j_4, j_3+j_4-1,...,|j_3-j_4|$.
 We choose the z-axis along the vector of the initial momentum
 $  {\hat {\bf  k}}$, therefore 
$Y_{Lm}({\hat {\bf k}})= \sqrt{(2\,L+1)/4\pi}\,\delta_{m\,0}$.
 Due to P-parity conservation,  the orbital
 momentum $L$ in Eq. (\ref{tfi}) is restricted by the condition
 $(-1)^L=\pi$, where $\pi=\pi_1\,\pi_2\,\pi_3\pi_4$ is the product
 of internal parities of the participating particles, $\pi_i$. 
 We consider here mainly  
 transitions without mixing the total isospin $T$ in this reaction
{\footnote {The isospin mixing is possible, for example, in the
 reaction $p+n\to \Sigma^0+\Theta^+$, if  the $\Theta^+$
is the isotriplet. In this case the P-parity can not be determined
by using the method in question.}.
 For the fixed $T$ and $\pi$
 the spin of the initial nucleons $S$ is fixed 
 unambiguously  by the generalized  Pauli principle:
 $(-1)^S=\pi (-1)^{T+1}$.  Therefore, in order to determine the
 P-parity $\pi$ of the system at given isospin $T$, it is sufficient 
 to determine  the spin of the NN-system 
 in the  initial state of this reaction.
 Using Eq.(\ref{tfi}) one can find the polarized cross section
 $d\sigma({\bf p}_1,{\bf p}_2)$  as the following
 \begin{eqnarray}
\label{polarsec}
d\sigma({\bf p}_1,{\bf p}_2)=\Phi \sum_{\mu_3\,\mu_4}
|T_{\mu_1\,\mu_2}^{\mu_3\,\mu_4}|^2=\frac{1}{4\pi} \sum_{M}
(\frac{1}{2}\mu_1\frac{1}{2}\mu_2|SM)^2\times \nonumber \\
 \times \sum_{J\,M\,L\,L'}
\sqrt{(2L+1)(2L'+1)}(S\,M\,L\,0|J\,M)(S\,M\,L'\,0|J\,M)
 a_J^{LS}\, (a_J^{L'S})^*,
\end{eqnarray} 
where $\Phi$ is a kinematical factor. 
 Using the relations $(\frac{1}{2}\mu_1\frac{1}{2}\mu_2|00)= 
 \chi_{\mu_1}^{+} \frac {i\sigma_y}{\sqrt{2}}\chi_{\mu_2}^{(T)+}$ and    
$(\frac{1}{2}\mu_1\frac{1}{2}\mu_2|1\lambda)= \chi_{\mu_1}^+\sigma_\lambda \frac {i\sigma_y}{\sqrt{2}}\chi_{\mu_2}^{(T)+}$, where
 $\sigma_i\, (i=y,\lambda$) is the Pauli matrix and $\chi_\mu$ is the
 2-spinor, one can find
\begin{equation}
\label{s0}
(\frac{1}{2}\mu_1\frac{1}{2}\mu_2|00)^2=
\frac{1}{4}(1-{\bf p}_1\cdot {\bf p}_2), 
 \end{equation}
\begin{equation}
\label{s1}
(\frac{1}{2}\mu_1\frac{1}{2}\mu_2|1M)^2=
\begin{cases}
\frac{1}{4}(1+{\bf
    p}_1\cdot {\bf p}_2-2p_{1z}p_{2z}),  {\text{  $M=0$,}} \\
\frac{1}{4}[1 \pm (p_{1z}+p_{2z})
    +p_{1z}p_{2z}], \ {\text {$M=\pm 1$}},
\end{cases} 
\end{equation}
In Eqs. (\ref{polarsec}), (\ref{s0}) and (\ref{s1}) 
   ${\bf p}_i$ is the polarization
 vector of the i-th particle with the spin $j_i=\frac{1}{2}$
 being in the pure spin state $\chi_{\mu_i}$. 
 The unpolarized cross section is given as
\begin{equation}
\label{unpols}
d\sigma_0= \Phi\,\frac{1}{4} \sum_{\mu_1\,\mu_2 \,\mu_3\,\mu_4}
|T_{\mu_1\,\mu_2}^{\mu_3\,\mu_4}|^2= 
 \frac{1}{16\,\pi} \Phi \,\sum_{J,L} (2J+1)|a_J^{L\,S}|^2.
\end{equation}

 \subsection  {The spin singlet initial state}   

 Using Eqs. (\ref{polarsec}), (\ref{s0}) and (\ref{unpols})
 one can find for the spin-singlet polarized cross section
 the following formula
 \begin{equation}
 \label{s0sec}
 d\sigma({\bf p}_1,{\bf p}_2)=d\sigma_0 (1-{\bf p}_1\cdot {\bf p}_2), 
 \end{equation}
   In notations of Ref.\cite{ohlsen}, non-zero spin-spin correlation
 parameters for this case are the following: $C_{x,x}=C_{y,y}=C_{z,z}=-1$.
 
 In order to find spin-transfer coefficients, one should consider
 the following cross section
 \begin{equation}
 \label{s0kij}
  d\sigma({\bf p}_1,{\bf p}_3)= \Phi\sum_{\mu_2,\mu_4}
|T_{\mu_1\,\mu_2}^{\mu_3\,\mu_4}|^2.
  \end{equation}
 The polarization vector ${\bf p}_1$ of the particle 1 in
 the right side of Eq.(\ref{s0kij}) can be found only
 in the following sum
  \begin{equation}
 \label{s0kij2}
 \sum_{\mu_2}(\frac{1}{2}\mu_1\frac{1}{2}\mu_2|00)^2=
\frac{1}{2}\sum_{\mu_2} (\chi_{\mu_1}^{+} {i\sigma_y}
 \chi_{\mu_2}^{(T)+})(\chi_{\mu_2}^{T}{(-i\sigma_y)}
\chi_{\mu_1})= 
\frac{1}{4} Sp(1+{\bfg \sigma} \cdot
 {\bf p}_1)=\frac{1}{2}.
 \end{equation}
 Since the vector ${\bf p}_1$ is absent actually in the right hand
 side of Eq.
 (\ref{s0kij2}), one should conclude that the all polarization 
 transfer coefficients are zero for the spin-singlet initial state:
 $K_i^j=0\, (i,j=x,y,z)$.  The obtained results for $C_{i,j}$ and $K_i^j$
 are valid for any values of the spins  $j_3$ and $j_4$.
 
 \subsection  {The spin triplet initial state}

  For $S=1$ and $ M=0$  Eq.(\ref{polarsec})   can be written as
 \begin{eqnarray}
 \label{m0}
 d\sigma^{M=0}({\bf p}_1,{\bf p}_2)=
\frac{\Phi}{16\,\pi}(1+{\bf
    p}_1\cdot {\bf p}_2-2p_{1z}p_{2z})\,\sum_J|\sqrt{J}\,
  a_J^{J-1}-\sqrt{J+1}\,a_J^{J+1}|^2. \ \ \ \ \ \ 
\end{eqnarray}
 We come to this formula from Eq.(\ref{polarsec})   using  
Eq. (\ref{s1}) and 
the  following formulae for the Clebsh-Gordan coefficients:
$(1\,0\, J\,0|J\,0)=0,(1\,0\, J-1\,0|J\,0)=\sqrt{J/(2J-1)}$, 
$(1\,0\, J+1\,0|J\,0)=\sqrt{J+1/(2J+3)}$.
 In order to  simplify the notations,  we omit in Eq.(\ref{m0})
 and below
  the superscript $S=1$ in $a_J^{L\,S}$. The sum over the projections
  $M= +1$ and $M=-1$ into right hand side of  Eq. (\ref{polarsec})
gives 
\begin{equation}
\label{mpm1}
 d\sigma^{M=\pm1}({\bf p}_1,{\bf p}_2)=
\frac{\Phi}{16\,\pi}(1+p_{1z}p_{2z})
\begin{cases}
\sum_J\,
|\sqrt{J}\, a_J^{J+1}+\sqrt{J+1}\,a_J^{J-1}|^2, &
 {\text {if \, $(-1)^J=\pi$,}}\\
\sum_{J}\,(2J+1)\,|a_J^J|^2, & {\text {if\, $(-1)^{J+1}=\pi$}},
\end{cases}
\end{equation}
 Here we used the following relations: $ (11\,J\,-1|J\,0)=\frac{1}{\sqrt{2}}$,
$(11\,J\,-1|J-1\,0)=\sqrt{J+1}/\sqrt{2(2J+1)}$,
$(11\,J\,-1|J+1\,0)=\sqrt{J}/\sqrt{2(2J+1)}$.
Using Eqs.(\ref{m0}), (\ref{mpm1}) and (\ref{unpols}), one can present
the polarized cross section (\ref{polarsec})
 in the  following standard form \cite{ohlsen} 
\begin{equation}
\label{standard}
 d\sigma({\bf p}_1,{\bf p}_2)= d\sigma_0\,
(1 + C_{x,x}\,p_{1x}\,p_{2x}+ C_{y,y}\,p_{1y}p_{2y}+
C_{z,z}\,p_{1z}\,p_{2z}),
\end{equation}
where the spin-spin correlation parameters given as
\begin{eqnarray}
\label{cyys1}
  C_{x,x}=C_{y,y}= \frac{\sum_J|\sqrt{J}\,
  a_J^{J-1}-\sqrt{J+1}\,a_J^{J+1}|^2} {\sum_{J\,L}(2J+1)\,|a_J^L|^2},
  \\
\label{czzs1}
 C_{z,z}=1-2\,C_{y,y},
\end{eqnarray}
As it seen from Eq. (\ref{cyys1}), the spin-spin correlation parameters
are non-negative for transversal polarization. 

 Considering the sum
 $\sum_{\mu_2}(\frac{1}{2}\mu_1\frac{1}{2}\mu_2|1M)
(\frac{1}{2}\mu_1\frac{1}{2}\mu_2|1M')$, one can find that
this sum contains explicitly  the polarization
 vector ${\bf p}_1$. Therefore, in contrast to the case
 of $S=0$, the spin-triplet initial state $S=1$ allows 
 a  non-zero polarization transfer in this reaction. 
 In order to get  the spin-transfer coefficients we use below
 a general method developed in Ref.\cite{bilenky}.

\section { General method} 

 According to Ref.\cite{bilenky},
  the amplitude in Eq.(\ref{tfi}) can be written as 
\begin{equation}
\label{operatortfi}
 T_{\mu_1\,\mu_2}^{\mu_3\,\mu_4}=
\chi_{j_3\,\mu_3}^+\chi_{j_4\,\mu_4}^+ {\hat F}\chi_{j_1\,\mu_1}\,
\chi_{j_2\,\mu_2},
\end{equation}
where $ {\hat F}$ is an operator acting on the spin-states  of the
initial and final particles.
 This operator can be written as
\begin{equation}
\label{f}
{\hat F}= \sum_{m_1\,m_2\,m_3\,m_4}
T_{m_1\,m_2}^{m_3\,m_4}\, \chi_{j_1\,m_1}^+(1) \chi_{j_2\,m_2}^+(2)
\chi_{j_3\,m_3}(3)\,\chi_{j_4\,m_4}(4),
\end{equation} 
where $\chi_{j_k\, m_k} (k)$ is the spin function of the $k$-th
particle with the spin $j_k$ and z-projection $m_k$ and 
$ T_{m_1\,m_2}^{m_3\,m_4}$  is defined by Eq. (\ref{tfi}).
The operator ${\hat F}$ is
 normalized to the unpolarized cross section as
\begin{equation}
\label{trff}
d\sigma_0 =\frac{\Phi}{(2j_1+1)\,(2j_2+1)} Sp FF^+.
\end{equation}

\subsection {Polarizaton transfer coefficients}

  The spin-transfer coefficient is given by the following formula
\cite{bilenky}
\begin{equation}
\label{kijff}
 K_\lambda^\kappa= \frac{Sp F\sigma_\lambda(1)\,F^+\,\sigma_\kappa(3)}
{Sp F\,F^+},
 \end{equation}
 where $\lambda,\kappa=0,\,\pm 1$.
 For  $j_1=j_3=\frac{1}{2}$ we found from
Eqs.(\ref{tfi}), (\ref{operatortfi}) and (\ref{kijff}) the spin transfer
coefficient in the following general form
\begin{eqnarray}
\label{kijgen}
4\,d\sigma_0 K_\lambda^{\kappa}=
\delta_{\lambda,-\kappa}\,\frac{3}{2\pi}
\sum_{S\,S'\,J\,J'\,L\,L'\, J_0}
\sqrt{(2L+1)\,(2L'+1)\,(2S+1)\,(2S'+1)}
 \times \nonumber \\
\times  (2J+1)\,(2J'+1)\,(-1)^{j_2+j_4 + S'+J'+L}
(1\, -\lambda \, 1\, \lambda |J_0\, 0)\,(L'\, 0 \, L\, 0\,|J_0 0) \times
\nonumber \\
\times \left \{\begin{array}{ccc}
\frac{1}{2} & j_2 & S \\
S' & 1 & \frac{1}{2}  
\end{array} \right \}
\left \{\begin{array}{ccc}
\frac{1}{2} & j_4 & J' \\
J & 1 & \frac{1}{2}  
\end{array} \right \}
\left \{\begin{array}{ccc}
J & S & L \\
J' & S' & L' \\
1 & 1 &  J_0 
\end{array} \right \} a_J^{L\,S} (a_{J'}^{L'\,S'})^* .\ \ \ \ \ \ \
\end{eqnarray} 
 Here we used the standard notations for the 6j-and 9j-symbols
 \cite{Varschalovich}. One can find from Eq.(\ref{kijgen}) the
 following relations
\begin{eqnarray}
\label{k}
K_{+1}^{-1}=K_{-1}^{+1}= -K_x^x=-K_y^y,
\end{eqnarray}
and $K_i^j=0$ at $i\not =j$, where $i,j=x,y,z$.
We find also from Eq.(\ref{kijgen}) that
 there is no polarization transfer 
 ($K_i^j=0$, $i,j=x,y,z)$ for $S=S'=0$ in accordance with above
 discussion. These coefficients equal to  zero also  for $J=J'=0$.
 For the spin-triplet transitions $S=S'=1$, we find from
 Eq. (\ref{kijgen}) that $K_x^x=K_y^y\not=0$ and  $K_0^0=K_z^z\not =0$.

 As an example, let us  consider the reaction with the minimal spins
 $j_i=\frac{1}{2},\, i=1,...,4$. For the total isospin $T=0$ and
 parity  $\pi=+1$
 one has got $S=1$. For this case Eq.(\ref{kijgen}) gives
 (using the notation $a_J^{L\,1}=a_J^L$)
\begin{eqnarray}
\label{onehalfplusx}
K_x^x=K_y^y=
\frac{|\sqrt{2}\,a_1^0+a_1^2|^2-3\,Re\,(\sqrt{2}\,a_1^0+a_1^2){a_1^2}^*}
{3\,(|a_1^0|^2+|a_1^2|^2)}, \\
\label{onehalfplusz}
K_z^z=
\frac{|\sqrt{2}\,a_1^0+a_1^2|^2}
{3\,(|a_1^0|^2+|a_1^2|^2)}. 
\end{eqnarray}
 The formulae (\ref{onehalfplusx}) and (\ref{onehalfplusz})
  coincide with those obtained previously in Ref.\cite{uzikov}
  by a different  method. For the case of  $T=1$
 $\pi=-1$ one has  got $S=1$. In this case  Eq.(\ref{kijgen})
 gives the following formulae
\begin{eqnarray}
\label{onehalfminusx}
 K_x^x=K_y^y=\frac{\sqrt{6}\, Re\,a_0^1\,{a_1^1}^*}
{|a_0^1|^2+3|a_1^1|^2},\\
 \label{onehalfminusz}
 K_z^z=\frac{{3}|a_1^1|^2}{|a_0^1|^2+3|a_1^1|^2}, 
\end{eqnarray}
 which  coincide (except for notations)  with those
 obtained recently in Ref. \cite{rekalo2} in the $\bfg
 \sigma$-representation for the amplitude. For higher  spins of the
 4-th particle $j_4\geq \frac{3}{2}$, Eq.(\ref{kijgen})
 also gives non-zero coefficients 
 $K_x^x$ and  $K_z^z$, but the formulae are  more cumbersome 
 and thus  we do not write  them here.

\subsection {Spin-spin correlation coefficients}
 
 For the spin-spin correlation coefficient, defined as \cite{ohlsen}
\begin{equation}
\label{clk}
 C_{\lambda\,\kappa}=\frac{Sp F\sigma_\lambda(1)\,\sigma_\kappa(2)\,F^+}
 {Sp F\,F^+},
 \end{equation}
we found for the case of
 $j_1=j_2=\frac{1}{2}$
\begin{eqnarray}
\label{cijgen}
4 d\sigma_0\,C_{\lambda,\kappa}=
\delta_{\lambda,-\kappa}\,
\frac{ 3}{2\pi}
\sum_{S\,S'\,J}(-1)^{S+J} (2J+1)
\, \sqrt{(2S+1)(2S'+1)}\times \nonumber \\
\times \sum_{L\,L\,J_0}(-1)^{L'}(2J_0+1)\sqrt{2L'+1}
(1\lambda \,1-\lambda\,|J_0\,0)\,(J_0\,0 \,L'\, 0\,|L\,0)\times \nonumber \\
\times  \left \{\begin{array}{ccc}
S' & S & J_0 \\
L & L' & J  
\end{array} \right \}
\left \{\begin{array}{ccc}
S^\prime & \frac{1}{2} & \frac{1}{2} \\
S & \frac{1}{2} &  \frac{1}{2}\\
J_0 & 1 &  1 
\end{array} \right \}
 a_J^{L\,S} (a_{J'}^{L'\,S'})^* .\ \ \ \ \ \ \
\end{eqnarray}
 We found from Eq. (\ref{cijgen}) the following relations:
$C_{+1,-1}=C_{-1,+1}=-C_{x,x}=-C_{y,y}\not =0$,  $C_0^0=C_z^z\not=0$,
whereas $C_{i,j}=0$ at $i\not =j$ ($i,j=x,y,z$). 

 One can see   from
 Eq.(\ref{clk}) that the sum  $\Sigma=C_{x,x}+C_{y,y}+C_{z,z}$
is equal to ${\bfg \sigma}(1)\cdot {\bfg\sigma}(2)$. Therefore
 $\Sigma$ is
 fixed by the spin $S$: $\Sigma=-3$
 for  $S=0$  and $\Sigma=+1$ for $S=1$
 in accordance with the above results given in 
 Eqs.(\ref{cyys1}), (\ref{czzs1}) and (\ref{s0sec}).
 One can find from 
 Eq.(\ref{cijgen}) that  $C_{x,x}=C_{y,y}=C_{z,z}=-1$ for $S=S'=0$.
 For $S=S'=1$
 we did not find  here a transformation from  the rather cumbersome formula
 (\ref{cijgen}) to  more compact form of Eqs. (\ref{cyys1}) and
 (\ref{czzs1}). However, one  can check
 by straightforward calculations that  these formulae
 lead to  the  same result for the spin-triplet initial state.

\section{ Full spin structure for the reaction $\frac{1}{2}+\frac{1}{2}\to
\frac{1}{2}+\frac{1}{2}$}

 For completeness, in this section we give the full spin structure 
 of the binary  reaction  at $j_1=j_2=j_3=j_4= \frac{1}{2}$
 discussed in part in Ref.\cite{uzikov}. 
 For the case $T=0$ and $\pi=-1$ one has got $S=0$, and
 the amplitude (\ref{tfi}) can be written as
\begin{equation}
\label{tfipn0}
 M_{\mu_1\,\mu_2}^{\mu_3\,\mu_4}=
 \sum_{\alpha=x,y,z} \,(\chi_{\mu_3}^+\sigma_\alpha\frac{i\sigma_y}{\sqrt{2}}\,
\chi_{\mu_4}^{(T)+})\,(\chi_{\mu_1}^{(T)}\frac{-i\sigma_y}{\sqrt{2}}\,
\chi_{\mu_2}) {\hat k}_\alpha \sqrt{\frac{3}{4\,\pi}}\, a_1^{1\,0}.
\end{equation}
When deriving Eq.(\ref{tfipn0}) from Eq.(\ref{tfi}) we used
 for the Clebsh-Gordan coefficients   
 the formulae given  above after Eq.(\ref{polarsec}).
  The unpolarized cross section corresponding
 to the amplitude (\ref{tfipn0}) takes the 
 following form \begin{equation}
\label{pn0unpol}
 d\sigma_0=\frac{1}{4}\Phi\,\sum_{\mu_1\,\mu_2\,\mu_3\,\mu_4}
 |M_{\mu_1\,\mu_2}^{\mu_3\,\mu_4}|^2=
 \frac{3}{16\pi}\Phi\,|a_1^{1\,0}|^2
 \end{equation}
 that is in agreement with Eq.(\ref{unpols}).
 In order to calculate the polarized cross section we use the 
 density matrix for the spin-$\frac{1}{2}$ particle being
 in the pure spin state $\chi_{\mu_i}$ in the following form
\begin{equation}
\label{density}
\chi_{\mu_i}\,\chi_{\mu_i}^+ =\frac{1}{2}(1+{\bfg \sigma}\cdot {\bf
  p}_i).
\end{equation}
 Using Eqs.(\ref{density}) and (\ref{tfipn0}) one can write the
  cross section with polarized  both initial and final particles as
\begin{eqnarray}
\label{fullpn0}
d\sigma({\bf p}_1, {\bf p}_2;{\bf p}_3, {\bf p}_4) =
\Phi\, |M_{\mu_1\,\mu_2}^{\mu_3\,\mu_4}|^2=\nonumber \\
=\frac{1}{4} d\sigma_0\, (1-{\bf p}_1\cdot {\bf p}_2)\,[ 1+ {\bf p}_3
\cdot {\bf p}_4-
 2({\bf p}_3\cdot {\hat {\bf k}})({\bf p}_4\cdot {\hat {\bf k}})].  
\end{eqnarray}
 The polarization vectors of the final particles ${\bf p}_3$ 
 and ${\bf p}_4$ are determined by the reaction amplitude
 (\ref{tfipn0}) and can be found using the standard methods
 \cite{bilenky,ohlsen}. After performing  this step and substituting
 obtained vectors ${\bf p}_3$  and ${\bf p}_4$ 
 into Eq.(\ref{fullpn0}), one can find   the polarized cross section
 $d\sigma({\bf p}_1, {\bf p}_2)$ given by  Eq.(\ref{s0sec}).
 However, the calculation of ${\bf p}_3$ and  ${\bf p}_4$ 
 is not necessarily  and Eq.(\ref{fullpn0}) is 
 sufficient to find all spin
 observables for the reaction described by the amplitude (\ref{tfipn0}).  
 In particular, one can see from Eq.(\ref{fullpn0}) that there is no
  polarization transfer in this reaction ($K_i^j=0,\,
 i,j=x,y,z$), but   there are  spin-spin correlations both  in
 the initial and  final  states.

 For  $T=0$ and  $\pi=+1$  we have got $S=1$,
 and the amplitude in Eq.(\ref{tfi}) can be
 written as
\begin{equation}
\label{tfipn1}
M_{\mu_1\,\mu_2}^{\mu_3\,\mu_4}
= \sum_{\alpha=x,y,z} \,(\chi_{\mu_3}^+\sigma_\alpha\frac{i\sigma_y}
{\sqrt{2}}\,
\chi_{\mu_4}^{(T)+})\,(\chi_{\mu_1}^{(T)}\frac{-i\sigma_y}{\sqrt{2}}\,
\Pi_\alpha \chi_{\mu_2}) \sqrt{\frac{3}{4\,\pi}}, 
\end{equation}
where $\Pi_\alpha$ is the following spin operator
\begin{equation}
\label{palpha}
\Pi_\alpha= G\sigma_\alpha +F\,
 {\hat {k}}_\alpha (\bfg \sigma \cdot {\hat {\bf k}})
\end{equation}
with
\begin{equation}
G=\frac{1}{\sqrt{4\pi}}\,(a_1^0+\frac{1}{\sqrt{2}}\, a_1^2)
\end{equation}
and
\begin{equation}
\label{f1}
F=-\frac{3}{\sqrt{8\pi}}\, a_1^2.
\end{equation}
The cross section with polarized initial and final particles
 is the following
\begin{eqnarray}
\label{spsp}
d\sigma({\bf p}_1, {\bf p}_2;{\bf p}_3, {\bf p}_4) =
\Phi\, |M_{\mu_1\,\mu_2}^{\mu_3\,\mu_4}|^2=
\sum_{\alpha\, \beta=x,y,z}
 \frac{1}{8}Sp\{\sigma_\alpha\,(1-{\bfg \sigma} \cdot {\bf p}_4)\,
\sigma_\beta (1+{\bfg \sigma} \cdot {\bf p}_3)\}\times\nonumber \\ 
\times \frac{1}{8}Sp\{\Pi^+_\alpha\,(1+{\bfg \sigma} \cdot {\bf p}_2)\,
\Pi_\beta (1-{\bfg \sigma} \cdot {\bf p}_1) \}.\ \ \ \ \  
\end{eqnarray}
Calculating the traces in Eq. (\ref{spsp}), one can find finally
\begin{eqnarray}
\label{spsp}
d\sigma({\bf p}_1, {\bf p}_2; {\bf p}_3, {\bf p}_4) =
\frac{1}{16}\Phi\, \bigl \{
|G|^2(1+{\bf p}_1\cdot {\bf p}_2)\, (3+{\bf p}_3\cdot {\bf p}_4)+\nonumber \\
+\bigl [(|F|^2+2 ReFG^*)(1+{\bf p}_1\cdot {\bf p}_2)\,
-2|F|^2\,({\bf p}_1\cdot {\hat {\bf k}})({\bf p}_2\cdot {\hat {\bf
    k}})\bigr ]
[1-2({\bf p}_3\cdot {\hat {\bf k}})({\bf p}_4\cdot {\hat {\bf k}})+{\bf p}_3
\cdot {\bf p}_4]-\nonumber \\
-2|G|^2\bigl [({\bf p}_1\cdot {\bf p}_2)(1+{\bf p}_3
\cdot {\bf p}_4)-({\bf p}_1\cdot {\bf p}_3)({\bf p}_2\cdot {\bf p}_4)
-({\bf p}_2\cdot {\bf p}_3)({\bf p}_1\cdot {\bf p}_4)
 \bigr ]-\nonumber \\
-2Re GF^* ({\bf p}_2\cdot {\hat {\bf k})\bigl [(\bf p}_1\cdot {\hat {\bf k})
(1+{\bf p}_3\cdot {\bf p}_4)-
({\bf p}_3\cdot {\hat {\bf k}})({\bf p}_4\cdot {\bf p}_1})
-({\bf p}_4\cdot {\hat {\bf k}})({\bf p}_1\cdot {\bf p}_3)\bigr ]-
\nonumber \\
-2Re GF^* ({\bf p}_1\cdot {\hat {\bf k})\bigl [(\bf p}_2\cdot {\hat {\bf k})
(1+{\bf p}_3\cdot {\bf p}_4)-
({\bf p}_3\cdot {\hat {\bf k}})({\bf p}_4\cdot {\bf p}_2})
-({\bf p}_4\cdot {\hat {\bf k}})({\bf p}_2\cdot {\bf p}_3)\bigr ]-
\nonumber \\
-2\,Im FG^*\, 
\bigl ([{\bf p}_1\times {\hat {\bf k}}]({\bf p}_2\cdot {\hat {\bf k})\
+(\bf p}_1\cdot {\hat {\bf k}})\, [{\bf p}_2\times {\hat {\bf
      k}}] \bigr )
\cdot ({\bf p}_3+{\bf p}_4)+ \nonumber \\
+2ImFG^*\,({\bf p}_3\cdot {\hat {\bf k}})
([{\hat {\bf k}}\times({\bf p}_1+{\bf p}_2)]
\cdot {\bf p}_4)+2ImFG^*\,({\bf p}_4\cdot {\hat {\bf k}})
([{\hat {\bf k}}\times({\bf p}_1+{\bf p}_2)]
\cdot {\bf p}_3)+
\nonumber \\
+2(|G|^2+Re\, FG^*)({\bf p}_1+{\bf p}_2)\cdot ({\bf p}_3+{\bf p}_4)
-2Re\, FG^*({\bf p}_1\cdot {\hat {\bf k}}+{\bf p}_2\cdot {\hat {\bf k}})
({\bf p}_3\cdot {\hat {\bf k}}+{\bf p}_4\cdot {\hat {\bf k}})\bigr
\}. \ \ \ \ \ \ \ \ 
\end{eqnarray}
 The unpolarized cross section for this case is
\begin{equation}
\label{pn1unpol}
 d\sigma_0= \frac{1}{4}\Phi\, \left \{
 |G+F|^2+2|G|^2 \right \} = \frac {\Phi}{16\pi}
 3(|a_1^0|^2+|a_1^2|^2).\ \ \ \ \ \ \ \ \
\end{equation}
Using Eq. (\ref{pn1unpol}), one can find  from Eq.(\ref{spsp})
 all spin observables for this reaction.
For example, one can see that  the spin-spin correlation coefficients
 $C_{i,j}$ and spin transfer coefficients $K_i^j$ obtained 
 from Eq.(\ref{spsp})   coincide  with those given by
 Eqs.(\ref{cyys1},\ref{czzs1}) and 
 Eqs.(\ref{onehalfplusx},\ref{onehalfplusz}), respectively.

\section{Conclusion}

  The obtained  formulae
 (\ref{s0sec}), (\ref{cyys1}), (\ref{czzs1}),
  (\ref{kijgen}) and (\ref{cijgen}) 
  allow us to conclude that for $S=1$
   (i) the spin-spin correlation coefficient
  $C_{y,y}$ is always non-negative, 
  and (ii) spin transfer coefficients $ K_y^y $ and $K_z^z$ are
  non-zero in the reaction in question  $1+2\to 3+4$ 
  at the  threshold  independently on the spin $j_4$ of the 4-th particle.
  On the contrary, for $S=0$, the spin-spin correlation coefficients
  $C_{x,x}=C_{y,y}=C_{z,z}$ equal to $-1$
  and all the spin transfer coefficients equal to zero.
   This conclusion  is a generalization of the previous results
 \cite{uzikov,rekalo2} found  for the case of $j_4=\frac{1}{2}$.
  The obtained  result
  allows one to determine unambiguously the P-parity of the $\Theta^+$
  by measurement of either  $C_{y,y}$ or $K_x^x$ (or $ K_z^z$)
  in the reaction
  $pp\to \Sigma^+\,\Theta^+$. The   total isospin of this channel is
  fixed, $T=1$, therefore the spin $S$ of the initial nucleons is directly
  related to the  P-parity $\pi_\Theta$ of the $\Theta^+$:
  $(-1)^S=\pi_\Theta$. In the reaction
  $pn\to \Lambda^0 \Theta^+$ one has got either  $(-1)^S=-\pi_\Theta$,
  if the isospin
  of the $\Theta^+$ is even ($I_\Theta=0$, $2$), or
  $(-1)^S=\pi_\Theta$, if $I_\Theta=1$.
  Therefore, both the P-parity and the isospin of the $\Theta^+$ 
  can be determined unambiguously  by combined measurement of
   $C_{y,y}$ or $K_y^y$ (or  $K_z^z$) in these two reactions.

{\bf Acknowledgment}. I am  thankful  to S.N.~Dymov, A.E.~Dorokhov,
 S.B.~Gerasimov,  V.I.~Komarov,
 H.~Str\"oher and O.P.~Teryaev  for  stimulating discussions.     


}
\end{document}